\documentstyle[prd,aps,preprint,tighten,epsfig]{revtex}
\begin{document}
\draft
\title{Timelike and spacelike QCD characteristics of the
$e^+e^-$ annihilation process}

\author{K.A. Milton\thanks{E-mail: milton@mail.nhn.ou.edu}}

\address{Department of Physics and Astronomy,
University of Oklahoma, Norman, OK 73019 USA}

\author{I.L. Solovtsov\thanks{E-mail: solovtso@thsun1.jinr.ru} and
O.P. Solovtsova}
\address{Bogoliubov Laboratory of Theoretical Physics,
Joint Institute for Nuclear Research, 141980 Dubna, Moscow Region, Russia}
\preprint{OKHEP--99--07}
\date{\today}
\maketitle
\begin{abstract}
We consider the Adler $D$-function, which is defined in the spacelike
region, and the $e^+e^-$annihilation ratio smeared, according to the Poggio,
Quinn, and  Weinberg method, which is  determined for timelike argument. We
show that the method of the nonperturbative $a$-expansion allows one to
describe these Euclidean and Minkowskian characteristics of the process of
$e^+e^-$ annihilation down to the lowest energy scales.
\end{abstract}

\vspace{0.7cm}
{\bf 1.} In quantum chromodynamics, it is important to determine the
``simplest" objects which allow one to check direct consequences of the
theory without using model assumptions in an essential manner.
Comparison of theoretical
results for these objects with experimental data allows us to
justify transparently the validity of basic statements of the theory,
and make some
conclusions about completeness and efficiency of the theoretical methods used.
Some single-argument functions which have a straightforward connection with
experimentally measured quantities can play the role of these objects.
A theoretical description of inclusive processes can be expressed in
terms of functions of this sort. Let us mention among them moments
${\rm{M}}_n(Q^2)$ of the structure functions in inelastic lepton-hadron
scattering and the hadronic correlator $\Pi(s)$ (or the corresponding Adler
$D$-function), which appear in the processes of $e^+e^-$ annihilation into
hadrons or the inclusive decay of the $\tau$ lepton.

In this paper we consider such objects for the process of $e^+e^-$
annihilation into hadrons. Here, the point is that the experimentally
measured ratio of hadronic and leptonic cross-sections, $R_{e^+e^-}(s)$, is
not suitable at the present stage of development of the theory to provide a
description  independent of model considerations. However, it is possible to
construct simpler objects than the $R_{e^+e^-}$-ratio. These are the Adler
function $D(Q^2)$~\cite{Adler}, which is defined in the Euclidean region,
and the quantity $R_\Delta(s)$ constructed by the Poggio, Quinn, and
Weinberg ``smearing" method~\cite{PQW} and defined in the Minkowskian
region.

The theoretical method, which we will use here, is the nonperturbative
expansion technique suggested in Refs.~\cite{Solovtsov94}. This approach is
based on the idea of variational perturbation theory (VPT)~\cite{VPT}, which
in the case of QCD leads to a new small expansion parameter. To compare the
results with experiment we will use the new ``experimental data'' for the
Adler function that has been recently obtained in Ref.~\cite{EJKV98} and the
smeared ``experimental curve'' corresponding to the $R$-ratio taken from
Refs.~\cite{MattinglyS94,BrodskyPT99}.

{\bf 2.} We use the method of constructing the so-called floating or
variational series in quantum theory. Within this approach, a certain
variational procedure is combined with the possibility of calculating
corrections to the principal contribution which allows the possibility of
probing the validity of the leading contribution and the region of
applicability of the results obtained. At present, this idea finds many
applications in the development of various approaches, which should enable us
to go beyond perturbation theory. Among these are the
Gaussian effective potential method~\cite{GEP}, the optimized
$\delta$-expansion~\cite{Delta-exp}, and the VPT approach~\cite{VPT}.

We will apply a nonperturbative QCD expansion based on a new small expansion
parameter~\cite{Solovtsov94}. Within this method, a quantity under
consideration can be represented in the form of a series, which is different
from the conventional
perturbative expansion and can be used to go beyond the
weak-coupling regime. This allows one to deal with considerably lower
energies than in the case of perturbation theory.

The new expansion parameter $a$ is connected with the initial coupling constant
$g$ by the relation
\begin{equation}
\label{lambda-a} \lambda\,=\,\frac{g^2}{{(4\pi)}^2}\,
=\,\frac{\alpha_s}{4\pi}\,=\,\frac{1}{C}\,\frac{a^2}{{(1-a)}^3}\, ,
\end{equation}
where $C$ is a positive constant. As follows from (\ref{lambda-a}), for any
value of the coupling constant $g$, the expansion parameter $a$ obeys the
inequality
\begin{equation} \label{a-ineq}
0 \leq a<1\, .
\end{equation}
While remaining within the range of applicability of the
 $a$-expan\-sion, one can
deal with low-energy processes where $g$ is no longer small.

The positive parameter $C$ plays the role of an auxiliary parameter of a
variational type, which is associated with the use of a floating series. The
original quantity, which is approximated by this expansion, does not depend
on the parameter $C$; however, any finite approximation depends on it due to
the truncation of the series. Here we will fix this parameter using some
further information,  coming from the potential approach to meson
spectroscopy. In the framework of this approach consider the following
approximations to the renormalization group $\beta$-function, the functions
${\beta}^{(3)}$ and ${\beta}^{(5)}$, which are obtained if one takes into
consideration the terms $O(a^3)$ and $O(a^5)$ in the corresponding
renormalization constant $Z_{\lambda}$. As has been shown in
Ref.~\cite{Solovtsov94}, $C$ is determined by requiring that
$-{\beta}^{(k)}(\lambda)/\lambda$ tends to 1 for sufficiently large
$\lambda$, which gives $C_3=4.1$ and $C_5=21.5$. The increase of $C_k$ with
the order of the expansion is explained by the necessity to compensate for
the higher order contributions. A similar phenomenon takes place also in
zero- and one-dimensional models. The behavior of the functions
$-\,{\beta}^{(k)}(\lambda)/{\lambda}$ gives evidence for the convergence of
the results, in accordance with the phenomenon of induced
convergence.\footnote{It has been observed empirically~\cite{Casw-K} that
the results seem to converge if the variational parameter is chosen, in each
order, according to some variational principle. This induced-convergence
phenomenon is also discussed in Ref.~\cite{Stevenson84}.} The behavior of
the $\beta$-function at large value of the coupling constant,
$-{\beta}^{(k)}(\lambda)/{\lambda}\simeq1$, corresponds to the infrared
singularity of the running coupling: ${\alpha}_s(Q^2)\sim Q^{-2}$ at small
$Q^2$. In the potential quark model this $Q^2$ behavior is associated with
the linear growth of the quark-anti\-quark potential.

The renormalization group $\beta$-function of the expansion parameter $a$ is
\begin{equation}
\label{beta-a} \beta_a(a)\,=\,\mu^2 \,\frac{\partial \,a}{\partial
\,\mu^2}\, =\,\frac{2\,\beta_0}{C}\,\frac{1}{F'(a)}\,,
\end{equation}
where $\beta_0=11-2f/3$ is the one-loop coefficient of the
$\beta$-function in the usual
perturbative expansion, and $f$ is the number of active
quarks, has a zero at $a=1$ that demonstrates the existence of the infrared
fixed point of the expansion parameter and its freezing-like behavior in the
infrared region. By finding the renormalization constants in the massless
renormalization scheme with an accuracy $O(a^3)$, we find
for the function $F(a)$
\begin{equation}
\label{f(3)} F^{(3)}(a)\,=\,\frac{2}{a^2}\,-\,\frac{6}{a}\, -48\,\ln\,a\,-
\,\frac{18}{11}\,\frac{1}{1-a}\,   +\,\frac{624}{121}\, \ln\,(1-a)\,
+\frac{5184}{121}\, \ln\,\left(1+\frac{9}{2}\, a\right)  \, .
\end{equation}

By solving the renormalization group equation (\ref{beta-a}) we find the
momentum dependence of the running expansion parameter $a(Q^2)$ as a
solution of the following transcendental equation
\begin{equation}
\label{RG-solution} \ln\,\frac{Q^2}{Q^2_0}\,=\,\frac{C}{2\,\beta_0}\,
\left[\,F(a)\,-\,F(a_0)\, \right] \, .
\end{equation}
For any values of $Q^2$, this equation has a unique solution $a=a(Q^2)$ in
the interval between $0$ and $1$.

By working at $O(a^5)$ we obtain a more complicated result
\begin{equation}\label{f(5)}
F^{(5)}(a) = \frac{1}{5(5+3B)}\,\sum_{i=1}^3 \, x_i\, J(a,b_i)
\end{equation}
with $B=\beta_1/(2C\beta_0)$, where the two-loop coefficient
$\beta_1=102-3f/3$, and
\begin{eqnarray}
J(a,b)& = &-\frac{2}{a^2b}-\frac{4}{ab^2}-\frac{12}{ab}
-\frac{9}{(1-a)(1-b)}+\frac{4+12b+21b^2}{b^3}\ln a \nonumber\\
&+&\frac{30-21b}{(1-b)^2}\ln(1-a)  -\frac{(2+b)^2}{b^3(1-b)^2}\ln(a-b)
\label{J(ab)}
\end{eqnarray}
with
\begin{equation}\label{x_i}
x_i  =  \frac{1}{(b_i-b_j)(b_i-b_k)}\, .
\end{equation}
Here indices $\{ijk\}$ are $\{123\}$ and cyclic permutations. The values of
$b_i$ are the roots of the equation $\psi(b_i)=0$, where the function
$\psi(a)$ is related to the $\beta$-function and is
\begin{equation}\label{psi(5)}
 \psi(a)  =  1+ \frac{9}{2}a + 2(6+a)a^2 + 5(5+3\, B)a^3\, .
\end{equation}
{\bf 3.} The cross-section for the process of $e^+e^-$ annihilation into
hadrons, or its ratio to the leptonic cross-section, $R(s)$, is a physically
measured quantity, defined for timelike momentum transfer -- the physical
region of the process. These quantities have a resonance structure that is
difficult to describe without model considerations. Moreover, the basic
method of performing calculations in quantum field theory, perturbation
theory, becomes ill-defined due to the so-called threshold singularities.
Both these problems can, in principle, be avoided if one considers a
``smeared" quantity. In the paper~\cite{PQW}, Poggio, Quinn, and Weinberg
suggested that instead of using the ratio
\begin{equation}
\label{R(s)}
R(s)=\frac{1}{2{\rm i}}
\left[\Pi(s+{\rm i}\epsilon)-\Pi(s-{\rm i}\epsilon)\right]\, ,
\end{equation}
defined through the hadronic correlation function $\Pi(q^2)$ taken near the cut,
which has a problem with threshold singularities, the following
smeared quantity be used:
\begin{equation}
\label{R(s)_Delta_1}
R_\Delta(s)=\frac{1}{2{\rm i}}
\left[\Pi(s+{\rm i}\Delta)-\Pi(s-{\rm i}\Delta)\right]
\end{equation}
with some finite $\Delta$. It has been argued in Ref.~\cite{PQW} that for
appropriate values of this smearing parameter ($\Delta$ is of order of a few
GeV$^2$) it is possible to compare theoretical predictions with smeared
experimental data. To get these data one can use the dispersion relation for
the hadronic correlator $\Pi(s)$ together with Eq.~(\ref{R(s)_Delta_1})
to give
\begin{equation}
\label{R(s)_Delta_2_1}
R_{\Delta}(s)=\frac{\Delta}{\pi} \int_0^\infty ds'
\frac{R(s')}{(s-s')^2+\Delta^2}\, .
\end{equation}
The corresponding ``experimental" curves of $R_{\Delta}(s)$ have been given
in Refs.~\cite{MattinglyS94,BrodskyPT99}. We will use these data to compare
our results with experiment.

However, a straightforward usage of conventional perturbation theory is not
still possible. Indeed, by parametrizing, as usual, the QCD contribution to
the function $R(s)$ in Eq.~(\ref{R(s)_Delta_2_1}) by the perturbative
running coupling, which has unphysical singularities, one encounters
difficulty with the  definition of the integral on the right-hand side.
Moreover, the usual method of the renormalization group gives a
$Q^2$-evolution law of the running coupling in the Euclidean region and
there is the question of how to parametrize, in terms of the same scale
parameter $\Lambda$, a quantity, for example $R(s)$, defined for timelike
momentum transfer. Here an important role is played by the analytic
properties of the running coupling. Within the nonperturbative $a$-expansion
it is possible to maintain such properties and self-consistently determine
the effective coupling in the Minkowskian region~\cite{JonesS95}. Note that
the so-called analytic approach to QCD~\cite{DVS96-97} also leads to a
well-defined procedure of analytic continuation from the spacelike to the
timelike domain~\cite{MS97,MS98}. By using the analytic approach the
characteristics of the $e^+e^-$ annihilation process have been analyzed in
Refs.~\cite{SS98,SS99}.

Another function which characterizes the process of $e^+e^-$ annihilation
into hadrons is the Adler function
\begin{equation}\label{D_def}
D(Q^2)=-Q^2\frac{d\,\Pi(-Q^2)}{d\,Q^2}
=Q^2\int_0^\infty ds\frac{R(s)}{(s+Q^2)^2}\,.
\end{equation}
Being defined outside of the resonance region, the $D$-function is the more useful
object to test QCD both in the ultraviolet perturbative region of large
$Q^2$ and in the nonperturbative region of small $Q^2$. Recently a new
``experimental" curve for this function has been obtained~\cite{EJKV98}.
This curve is a smooth function of $Q^2$ without any
traces of resonance structure, which makes it useful for  comparison with
reasonable theoretical descriptions. We will now consider $D(Q^2)$ and
$R_\Delta(s)$ in the framework of the nonperturbative $a$-expansion.

In the massless case, let us represent these functions in the form
\begin{equation}\label{D-lam_eff}
D(Q^2)=3\sum\limits_f q^2_f \left[1+d_0\lambda^{\rm eff}(Q^2) \right]
\end{equation}
and
\begin{equation}\label{R-lam_eff}
R(s)=3\sum\limits_f q^2_f \left[1+r_0\lambda^{\rm eff}_s(s) \right]\, .
\end{equation}
Here $q_f$ are the quark charges, $d_0$ and $r_0$ are the first perturbative
coefficients which are renormalization scheme independent, being
$d_0=r_0=4$. For the $D$-function defined in the spacelike region we can
write down the effective coupling in the form of the $a$-expansion which at
the $O(a^5)$ level has the form
\begin{equation}\label{lam_eff-a}
\lambda^{\rm eff}=\frac{1}{C}\,a^2+\frac{3}{C}\,a^3+
\left(\frac{6}{C}+\frac{1}{C^2}\frac{d_1}{d_0}\right)\,a^4
+\left(\frac{10}{C}+\frac{6}{C^2}\frac{d_1}{d_0}\right)\,a^5\, ,
\end{equation}
where $d_1$ is the next coefficient of the perturbative representation for
the $D$-function which we will take in the ${\overline{\rm MS}}$
renormalization scheme~\cite{GKL_NC86}
\begin{equation}
\label{d1-MS}
d_1=\frac{2}{3}\,\left[\,365-22f-8\zeta(3)\,(33-2f)\,\right]
\simeq 16\,(1.986-0.115f)\, .
\end{equation}
Here $\zeta(n)$ is the Riemann $\zeta$-function, $\zeta(3)\simeq1.202$.

Eq.~(\ref{R-lam_eff}) serves to define the effective coupling in the
timelike domain or as we will say in the $s$-channel, which is reflected in the
subscript $s$. For self-consistency of Eqs.~(\ref{D-lam_eff}) and
(\ref{R-lam_eff}) with the dispersion representation in Eq.~(\ref{D_def}) it
is important to maintain the correct analytic properties of the initial
effective coupling $\lambda^{\rm eff}(Q^2)$~\cite{JonesS95,MS97,MS98}. In this
case one finds
\begin{equation}
\lambda^{\rm eff}(Q^2)\,=\,Q^2\,\int_0^{\infty}\,
\frac{ds}{{(s+Q^2)}^2}\,\lambda_s^{\rm eff}(s)
\label{lambda_t-s}
\end{equation}
and the corresponding inverse relation
\begin{equation}
\label{lambda_s-t} \lambda_s^{\rm eff}(s)\,=\,-\,\frac{1}{2\pi {\rm i}}\,
\int _{s-{\rm i}\,\epsilon} ^{s+{\rm i}\,\epsilon} \frac{dz}{z}\,
\lambda^{\rm eff}(-z)\,=\,\frac{1}{2\pi {\rm i}}\,
\oint\limits_{|z|=s}\,\frac{dz}{z}\,\lambda^{\rm eff}(-z)\, .
\end{equation}

The $s$-channel running coupling can be written in the form
\begin{equation}
\label{lambda_fi} \lambda_s^{(i)}(s)\,=\,\frac{1}{2\pi{\rm
i}}\,\frac{1}{2\,\beta_0}\,\bigl[\,{\phi}^{(i)}(a_{+})\,
-\,{\phi}^{(i)}(a_{-})\,\bigr]\, ,
\end{equation}
where $a_{\pm}$ obey the equation
\begin{equation}
\label{f_pi} F(a_{\pm})\,=\,F(a_0)\,+\,\frac{2\,\beta_0}{C}\, \left[\,\ln\,
\frac{s}{Q_0^2}\,\pm\,{\rm i} \,\pi\,\right]\, .
\end{equation}
At the level $O(a^3)$ the function $\phi(a)$ has the form
\begin{equation}
\label{fi_3} {\phi}^{(3)}(a) =-4\ln a-\frac{72}{11}\frac{1}{1-a}+
\frac{318}{121} \ln (1-a)+\frac{256}{363} \ln \left(1+\frac{9}{2}a\right)\, .
\end{equation}
Similarly, a more complicated expression for the $O(a^5)$ level which we will
use can be derived. Note here that, as it has been recently argued from
general priciples, the behavior of the effective couplings in the spacelike and
the timelike regions cannot be symmetrical~\cite{MS-99}.

To incorporate quark mass effects, we use for the cross-section ratio an
approximate expression~\cite{PQW}
\begin{equation}
\label{R_approx} \tilde{R}(s)=3\sum_f\, q_f^2\, \Theta (s-4m_f^2)\, T(v_f)\,
\left[ 1\,+\,g(v_f) r_f(s) \right]\, ,
\end{equation}
 where the sum is performed over quark flavor and
\begin{eqnarray}
v_f&=&\sqrt{1-\frac{4m_f^2}{s}}\, , \qquad  T(v)=\frac{v(3-v^2)}{2}\, ,
\nonumber\\[0.3cm]
g(v)&=&\frac{4\pi}{3}\left[\frac{\pi}{2v}-\frac{3+v}{4}
\left(\frac{\pi}{2}-\frac{3}{4\pi} \right) \right]\, .
\label{vTg}\end{eqnarray}
The quantity $r_f(s)$ is defined by the $s$-channel effective coupling
$\lambda^{\rm eff}(s)$. The corresponding $D$-function can be found from
Eq.~(\ref{D_def}).

For massless, ${\rm MS}$-like, renormalization schemes one has to consider
some matching procedure. To this end, one usually adopts a procedure for
matching the running coupling in the Euclidean region~\cite{Marciano84} by
requiring that the running couplings,  corresponding to $(f-1)$ and $f$
number of fermions, should coincide with each other at some matching point
$Q=\xi\,m_f$. For the matching parameter $\xi$ one usually takes
$1\leq\xi\leq 2$. Obviously, in this case the derivative of the running
coupling will not be a continuous function and the correct analytic
properties of the $D$-function will be violated. Instead we will apply the
matching procedure in the $s$-channel, where, at least in the leading order,
the number of active quarks is directly connected with the energy $\sqrt{s}$
and discontinuously changes at the threshold $s=4m_f^2$.\footnote{In the
framework of the analytic approach, where  there is also a well-defined
procedure of analytic continuation, it is possible to use the same
$s$-channel matching method~\cite{MS98}.} The effective charge in the
Euclidean region, restored by the dispersion relation~(\ref{lambda_t-s}),
will have the correct analytic properties and will ``know," in a way similar
to massive MOM renormalization schemes, about all physical thresholds.

To perform this matching procedure, one can require that the $s$-channel running
coupling and its derivative be continuous functions in the vicinity
of the threshold. This requirement leads to a system of equations, which we
write down for a simple $O(a^3)$ case:
\begin{eqnarray}
\label{matching_sys}
&&\frac{1}{\beta_0^{f-1}}{\rm Im}\,\phi
\left(a_+^{f-1}\right)= \frac{1}{\beta_0^f}\,{\rm Im}\,\phi
\left(a_+^{f}\right)\, ,\\[0.3cm]
&&\frac{1}{C_{f-1}}{\rm Im}\,\left[ {\left(a_+^{f-1}\right)}^2
\left(1+3a_+^{f-1}\right)\right] =\frac{1}{C_{f}}{\rm Im}\,\left[\,
{\left(a_+^{f}\right)}^2 \left(1+3a_+^{f}\right)\right]\, . \nonumber
\end{eqnarray}
The function $\phi (a)$ is defined by Eq.~(\ref{fi_3}) and $a_+$ obeys to
Eq.~(\ref{f_pi}). Therefore, Eqs.~(\ref{matching_sys}) allows one to
establish  relations between the parameters $C_{f-1}\,$, $a_0^{f-1}$ and
$C_f\,$, $a_0^{f}$.

\begin{figure}[hbt]
\centerline{\epsfig{file=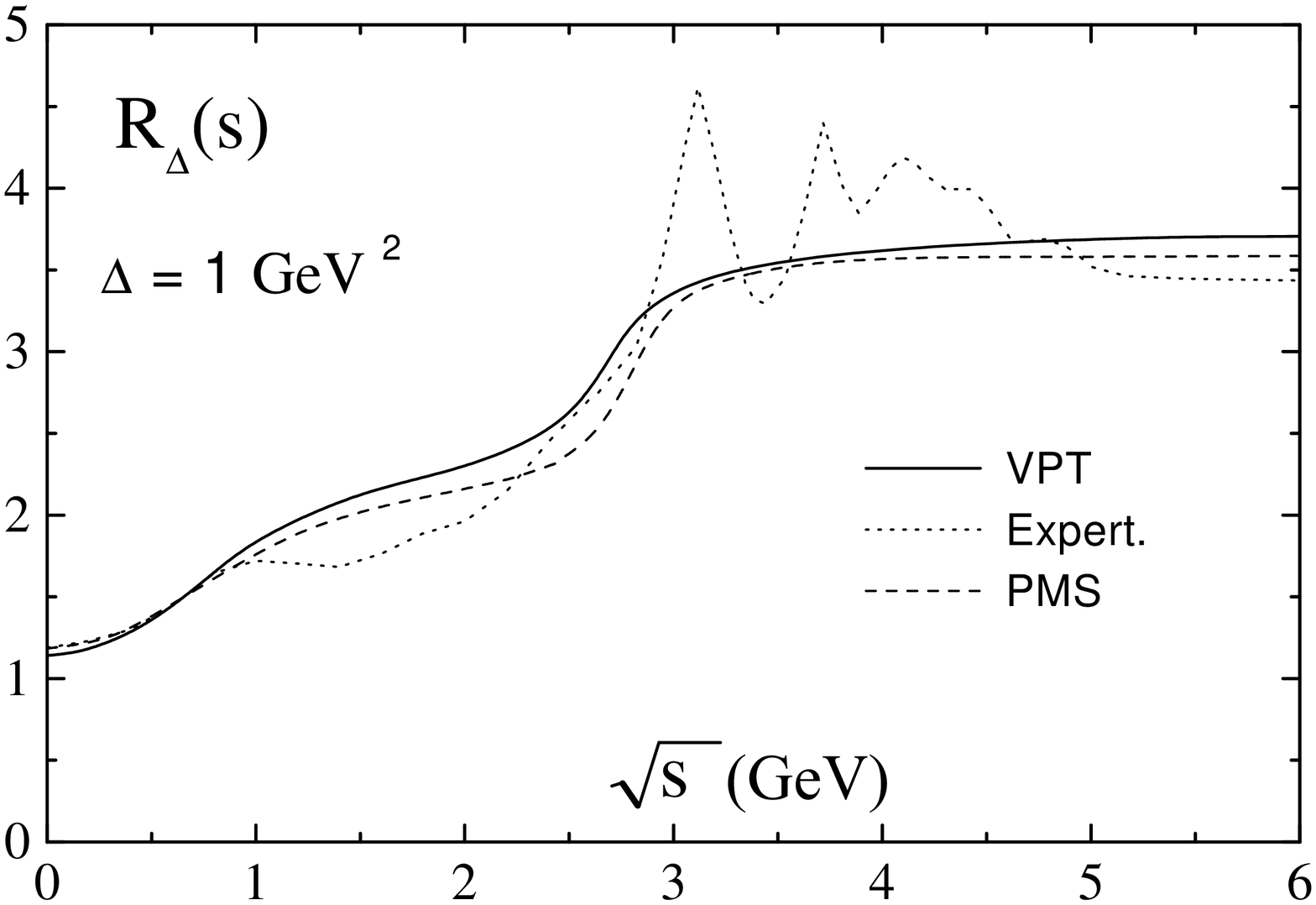,width=13cm}}
\caption{{\sl The smeared function $R_\Delta(s)$ for $\Delta=1~{\rm GeV}^2$.
The solid line corresponds to the VPT next-\-to-\-lea\-ding order (NLO)
result. The PMS result (the dashed curve) and the smeared experimental
curve (the dotted line) taken from Ref.~\protect\cite{MattinglyS94}.}}
         \label{R1_vpt}
         \end{figure}

\begin{figure}[hbt]
\centerline{\epsfig{file=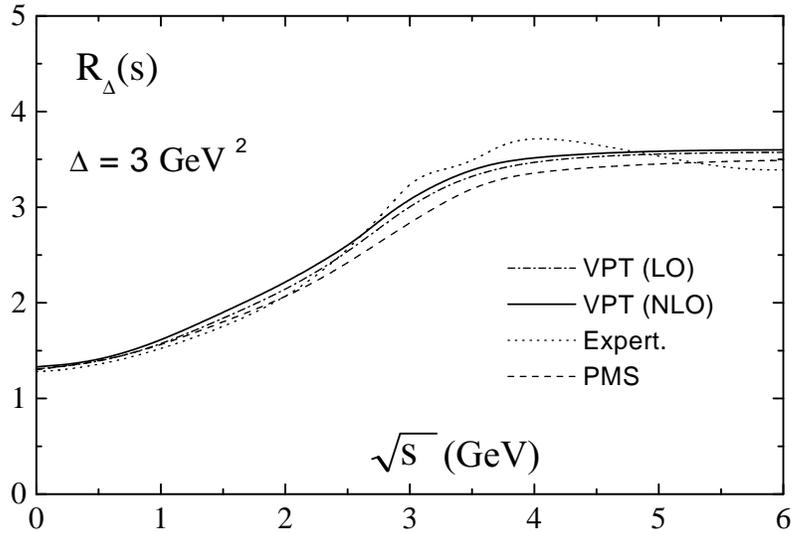,width=13cm}}
\caption{{\sl The smeared quantity $R_\Delta(s)$ for $\Delta=3~{\rm GeV}^2$.
The solid, dashed and dotted curves are defined as in Fig.~\protect\ref{R1_vpt}.
The dash-dotted curve corresponds to the leading VPT order.}}
         \label{R3_vpt}
         \end{figure}

\begin{figure}[hbt]
\centerline{\epsfig{file=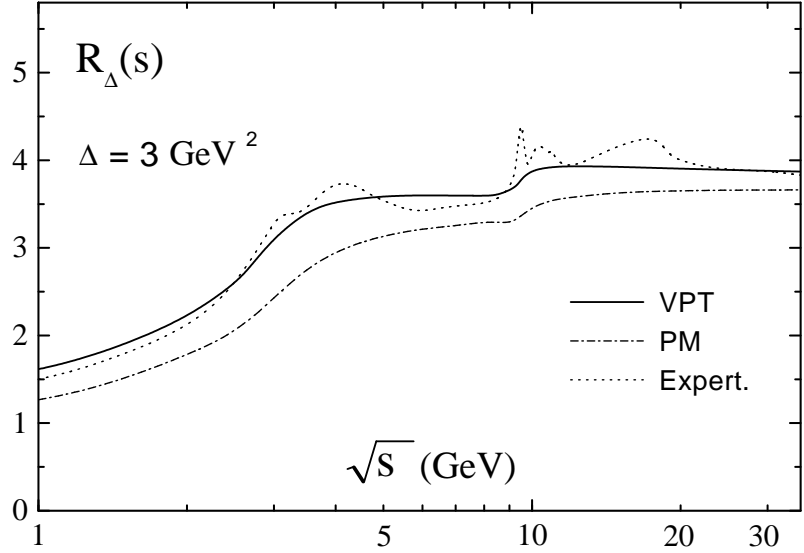,width=14cm}}
\caption{{\sl The smeared quantity $R_\Delta(s)$ for $\Delta=3~{\rm GeV}^2$.
The solid curve is the VPT result. The parton model (PM) prediction is
presented by the dash-dotted line and the experimental curve from
Ref.~\protect\cite{BrodskyPT99} is shown as the dotted line.}}
         \label{R3_Br_vpt}
         \end{figure}

\begin{figure}[hbt]
\centerline{\epsfig{file=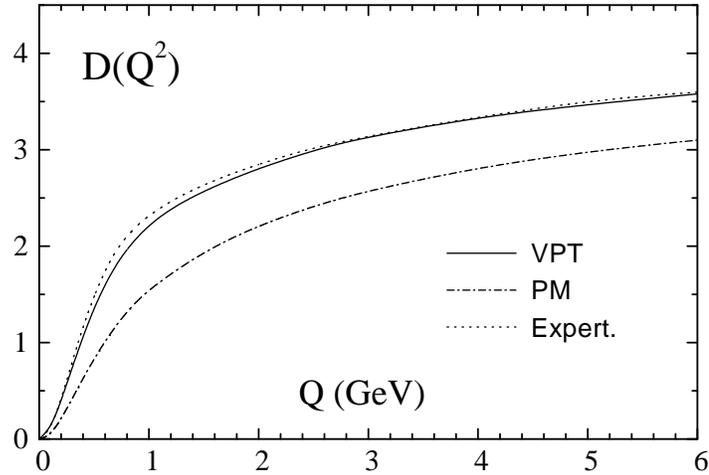,width=12.5cm}}
\caption{{\sl   The plot of the $D$-function.
The solid curve is the VPT result. The parton model (PM) prediction is
presented by the dash-dotted line and the experimental curve from
Ref.~\protect\cite{EJKV98} is shown as the dotted line.}}
         \label{D_vpt}
         \end{figure}

Our results are presented in Figs.~\ref{R1_vpt}--\ref{D_vpt}. In
Figs.~\ref{R1_vpt} and \ref{R3_vpt} we plot the smeared functions
$R_\Delta(s)$ for $\Delta=1~{\rm GeV}^2$ and $\Delta=3~{\rm GeV}^2$
respectively. The solid line is the VPT next-\-to-\-lea\-ding order (NLO)
result which was normalized at the $\tau$ lepton scale. To this end, we use
the method of description of the $\tau$ decay suggested in
Refs.~\cite{JonesSS95,JonesSS95_Brus} and take here the following
experimental average value $R_{\tau}=3.642\pm 0.021$~\cite{PDG98} as input.
To demonstrate the fact of stability we show in Fig.~\ref{R3_vpt} the
leading-order (LO) VPT result. In these figures we also plot, as dashed
curves, results obtained by using the principle of minimal sensitivity (PMS)
to optimize the third order of the perturbative expansion\footnote{The PMS
results correspond to a scale parameter
$\Lambda_{\overline{\rm{MS}}}^{(f=3)}=281$~MeV that is rather small as
compared with more modern value
$\Lambda_{\overline{\rm{MS}}}^{(f=3)}\simeq370$~MeV~\cite{ALEPH98}, which is
compatible with $R_\tau$-ratio used here.} and the smeared experimental
data, as dotted lines, taken from Ref.~\cite{MattinglyS94}. For
$\Delta=1~{\rm GeV}^2$ the resonances in the $e^+e^-$ annihilation
cross-section are not smeared enough, since there are some peaks in the
region of $J/\psi$ meson family. With increase in the value of $\Delta$ the
resonance structure is smoothed and finally disappears. For a wider interval
of energy up to 35 GeV the smeared experimental data have been recently
obtained in Ref.~\cite{BrodskyPT99}. We represent these data and our NLO
result in Fig.~\ref{R3_Br_vpt}, where we also show the parton model (PM)
result as the dash-dotted curve.

The $D$-function defined in the Euclidean region for positive momentum $Q^2$
is a smooth function and thus it is not necessary to apply any ``smearing"
procedure in order to have the possibility of comparing theoretical results
with experimental data. We plot our results in Fig.~\ref{D_vpt}, where we
also show the experimental curve taken from Ref.~\cite{EJKV98} and parton
model prediction. The shape of the infrared tail of the $D$-function is
sensitive to the value of the masses of the light quarks (the smeared
$R_\Delta(s)$ function for $\Delta\simeq1\mbox{--}3\;{\rm{GeV}}^2$ is less
sensitive to the value of these masses). In our calculations we use the
following masses $m_u=m_d=250\,{\rm{MeV}}$, $m_s=400\,{\rm MeV}$,
$m_c=1.3\,{\rm GeV}$, and $m_b=4.7\,{\rm GeV}$, which are close to the
constituent quark masses and incorporate, therefore, some nonperturbative
effects. Practically the same values of the quark masses were used to
describe the experimental data  in Refs.~\cite{SS99,Sanda}.

{\bf 4.} In this paper we have applied the nonperturbative method of the
$a$-expansion to describe the single-argument functions which are directly
connected with the experimental data describing $e^+e^-$ annihilation into
hadrons. An important feature of this approach is the fact that for
sufficiently small value of the coupling (the ultraviolet region of
momentum) it automatically reproduces the conventional perturbative results.
We emphasize that the entire high-energy physics regime is accessible within
this approach. Even going into the infrared region of small momentum, where
the running coupling becomes large and the standard perturbative expansion
fails, the $a$-expansion parameter remains small and we do not find
ourselves outside the region of validity of the approach.

We have considered two quantities that are convenient both for the
theoretical analysis and for the model-independent comparison with
experimental data. The first one is the Pog\-gio-\-Quinn-\-Wein\-berg
smeared function, $R_\Delta(s)$, defined in the Minkowskian region. The
second one is the Adler $D$-function defined in the Euclidean region. It
should be emphasized that the smeared quantity $R_\Delta(q^2)$ and the
$D$-function have different sensitivity to QCD parameters. For example, in
contrast to the smeared function taken with a reasonable value of the
parameter $\Delta$ the shape of the $D$-function in the infrared region is
very sensitive to the values of the masses of the light quarks. At the same
time, for $\Delta\simeq 2\mbox{--}3\,{\rm GeV}^2$, the function
$R_\Delta(s)$ is sensitive to the mass of charmed quark. Thus, these
functions can test different parameters of the theory and, in some sense,
are complementary to each other.

It should be stressed that in this approach we use the parameters which are
included in the Lagrangian and do not introduce any additional model
parameters. Nevertheless we found  good agreement between our results and
the experimental data down to the lowest energy scale. At the same time,
note that a straightforward attempt to apply the conventional operator
product expansion, which leads to the ``condensate'' term in the
$D$-function, a $\langle\alpha_sG^2\rangle/Q^4$ contribution, is not
satisfactory because of the ill-definition of this expansion at small
momentum.

\section*{Acknowledgement}
The authors would like to thank A.L.~Kataev, D.V.~Shirkov and A.N.~Sissakian
for useful discussions and interest in this work. Partial support of the
work by the US National Science Foundation, grant PHY-9600421,  by the U.S.
Department of Energy, Grant DE-\-FG-\-03-\-98ER41066, by the University of
Oklahoma, and by the RFBR, Grants 99-02-17727 and 99-01-00091, is gratefully
acknowledged.


\end{document}